\documentclass[preprint,superscriptaddress,nofootinbib]{revtex4-1} 
\pdfoutput=1

\usepackage{graphicx}  
\usepackage{dcolumn}   
\usepackage{bm}        
\usepackage{amssymb,amsmath}   
\usepackage{color}

\allowdisplaybreaks

\begin{document}


\title{Precise top-quark mass from the diphoton mass spectrum}

\author{Sayaka Kawabata}
\email{skawabata@seoultech.ac.kr}
\affiliation{Institute of Convergence Fundamental Studies, Seoul
National University of Science and Technology, Seoul 01811, Korea}

\author{Hiroshi Yokoya}
\email{hyokoya@kias.re.kr}
\affiliation{Quantum Universe Center, KIAS, Seoul 02455, Korea}


\begin{abstract}
We calculate the $gg\to\gamma\gamma$ amplitude by including the
 $t\bar t$ bound-state effects near their mass threshold.
In terms of the non-relativistic expansion of the amplitude, the LO
 contribution is an energy-independent term in the one-loop amplitude.
We include the NLO contribution described by the non-relativistic
 Green function and part of the NNLO contribution.
Despite a missing NLO piece which can be accomplished with the
 two-loop-level amplitude via massive quarks, the shape of the
 diphoton mass spectrum is predicted with a good accuracy.
Thanks to the simple and clean nature of the observable, its
 experimental measurement can be a direct method to determine the
 short-distance mass of the top quark at hadron colliders. 
\end{abstract}

\preprint{}

\maketitle

\section{Introduction}
 
At the LHC, a diphoton mass spectrum $d\sigma/dm_{\gamma\gamma}$
has attracted broad attentions for observations of the properties of the
Higgs boson in the standard model (SM)~\cite{Aad:2012tfa,%
Chatrchyan:2012xdj,Khachatryan:2014ira,Aad:2015tna,Khachatryan:2015rxa}
and searches for new phenomena beyond the SM~\cite{Aaboud:2016tru,%
Khachatryan:2016hje,Khachatryan:2015qba,Aad:2014ioa,%
ATLAS:2016eeo,Khachatryan:2016yec}.
At hadron colliders, pairs of high-$p_T$ photon are produced by $q\bar
q$ annihilation and gluon-fusion mechanisms~\cite{Catani:2011qz,%
deFlorian:2013psa,Campbell:2016yrh}, and processes
which involve fragmentation-photon contributions~\cite{Binoth:1999qq}. 
The $gg\to\gamma\gamma$ process, which is the main focus of this paper,
is described by loop diagrams with quarks in the SM. 
The analytic expression of the one-loop amplitude has been known for a
long time for both the massless- and massive-quark loops~\cite{%
Karplus:1950zz,DeTollis:1964una,DeTollis:1965vna,Costantini:1971cj,%
Combridge:1980sx,Dicus:1987fk,Bern:1995db}.
The two-loop amplitude has been calculated only for the
massless-quark loops~\cite{Bern:2001df,Bern:2001dg,Bern:2002jx}.

The threshold structure of the massive-quark-loop amplitude deserves 
particular interests~\cite{Dicus:1987fk,Chway:2015lzg,Jain:2016kai}
where the massive quark is regarded as the top quark or a hypothetical
particle beyond the SM.
Beyond the one-loop level, the amplitude receives large QCD corrections
due to the Coulomb-gluon exchanges between the nearly on-shell and
low-velocity heavy quarks in $s$-channel. 
Thus, the description of the amplitude requires an elaborate treatment
based on the non-relativistic QCD formalism.
For $gg\to\gamma\gamma$ process, such a study cannot be found in the
literature.
The aim of this paper is to compile the present knowledge of the
non-relativistic QCD theory for the description of the bound-state
effects in the massive-quark-loop amplitude, and to present a dedicated
and quantitative study on the diphoton mass spectrum near
the $t\bar t$ threshold.
Our framework follows the preceding studies on
$h\to\gamma\gamma$~\cite{Melnikov:1994jb,KY}, and some of our numerical
results overlap with that in Ref.~\cite{Chway:2015lzg}.

We discuss further to utilize the predicted mass spectrum for a precise 
determination of the top-quark mass, which is one of the fundamental
parameters in the SM.
Although the top-quark mass has been measured with an error of sub-GeV
level~\cite{Agashe:2014kda}, its interpretation in terms of well-defined
mass parameters is not settled yet in perturbative QCD.
It is known that the well-defined mass parameters can be determined by
using the threshold scan method at future $e^+e^-$
colliders~\cite{Martinez:2002st,Seidel:2013sqa,Horiguchi:2013wra}.
We show that the diphoton mass spectrum measurement can be a
considerable alternative to it at hadron colliders.
The application of the formula for physics beyond the SM will be
reported elsewhere.

\section{Scattering amplitude in the threshold limit}

We start the main content of the paper by introducing the scattering
amplitude for $gg\to\gamma\gamma$ at the one-loop level with the top
quark, and provide an easy-to-use expression for its threshold
behavior.
By using the all-outgoing convention for the momenta ($p_i$) and
helicities ($\lambda_i$), $g^{a_1}(-p_1,-\lambda_1) +
g^{a_2}(-p_2,-\lambda_2) \to \gamma(p_3,\lambda_3) +
\gamma(p_4,\lambda_4)$, the one-loop amplitude is written
as
\begin{align}
& {\mathcal M}_{gg\to\gamma\gamma} (\{p_i\};\{\lambda_i\};a_1,a_2) = 
 4\alpha\alpha_s \delta^{a_1a_2} \\
 & \times\left[ \left(\sum_{j=1}^{n_f}Q_j^2\right)
 M_{q,\{\lambda_i\}}(\{p_i\}) + Q_t^2 M_{t,\{\lambda_i\}}(\{p_i\};m_t)
 \right] \nonumber, 
\end{align}
where $M_q$ is the contribution from the massless-quark loop
with five flavors ($n_f=5$), and $M_t$ from the top-quark loop with
the top-quark pole-mass, $m_t$.
The amplitude for the top-quark loop near the threshold is expressed
as 
\begin{align}
 & M_{t,\{\lambda_i\}} = {\mathcal A}_{t,\{\lambda_i\}}(\theta) +
 {\mathcal B}_{t,\{\lambda_i\}} G^{(0)}(\vec0;E) + {\mathcal O}(v^2) ,
\end{align}
where $E \equiv m_{\gamma\gamma}-2m_t\simeq m_tv^2$ and
$v = \sqrt{1-4m_t^2/m^2_{\gamma\gamma}}$.
$G^{(0)}(\vec0;E) \equiv -m_t^2/(4\pi)\sqrt{-E/m_t-i\epsilon}$ is the
$t\bar t$ Green function in $S$-wave without QCD effects.
The first term which is energy-independent, represents the
contribution from the hard-momentum integral.
The second term which is ${\mathcal O}(v)$, represents the contribution
from the soft-momentum loop where the top-quarks can be on-shell.
At the one-loop level, all the imaginary part of the amplitude 
originates from $G^{(0)}$ above the threshold, $E\ge0$.
${\mathcal A}_t$ depends on the scattering angle $\theta$, while 
${\mathcal B}_t$ is independent of $\theta$ because only the
spin-singlet $t\bar t$ state contributes at this order. 
For $\{\lambda_i\}=\lambda_1\lambda_2\lambda_3\lambda_4$, we find
${\mathcal B}_{t,++++} = -{\mathcal B}_{t,--++} = -4\pi^2/m_t^2$, while
${\mathcal B}_{t,-+++} = {\mathcal B}_{t,-+-+} = 0$.
For the other combinations of the helicity, ${\mathcal B}_t$ as well as
${\mathcal A}_t$ can be written in terms of them.
For a description of ${\mathcal A}_t$, we make use of the partial-wave
decomposition with numerical coefficients.
The ${\mathcal A}_t$ term is expanded as
\begin{align}
 {\mathcal A}_{t,\{\lambda_i\}} (\theta) = \sum_{J=0}^{\infty}
 (2J+1){\mathcal A}^{J}_{t,\{\lambda_i\}} d^{J}_{\mu\mu'}(\theta) ,
\end{align}
where $\mu=-\lambda_1+\lambda_2$ and $\mu'=\lambda_3-\lambda_4$.
Because ${\mathcal B}_t$ is constant, ${\mathcal B}_t$ has only the
$J=0$ component, ${\mathcal B}_t={\mathcal B}_t^{J=0}$.
In Table~\ref{tab:AJ}, we list the numerical values of ${\mathcal
A}^{J}_{t}$ for $J$ up to 4.
The Wigner $d$-functions $d^{J}_{\mu\mu'}$ can be found in the
literature.
We find that the expansion up to $J=4$ gives a sufficiently good
approximation.

\begin{table*}[b]
\begin{tabular}{|c||c|c|c|c|c|}
 \hline
 $\lambda_1\lambda_2\lambda_3\lambda_4$ & ${\mathcal A}_t^{J=0}$ &
	 ${\mathcal A}_t^{J=1}$ & ${\mathcal A}_t^{J=2}$ & ${\mathcal
		 A}_t^{J=3}$ & ${\mathcal A}_t^{J=4}$ \\ 
 \hline
 \hline
 $++++$ & -1.06635650 & 0 & -0.00497776 & 0 & -0.00005389 \\
 \hline
 $--++$ & 1.57380190 & 0 & 0.00213711 & 0 & 0.00001690 \\
 \hline
 $-+++$ & - & - & -0.00290941 & 0 & -0.00001042 \\
 \hline
 $-++-$ & - & - & 0.11920027 & -0.00060737 & 0.00029467 \\
 \hline
\end{tabular}
 \caption{Numerical coefficients of ${\mathcal
 A}^{J}_{t,\lambda_1\lambda_2\lambda_3\lambda_4}$.}
 \label{tab:AJ}
\end{table*}
%

\section{Threshold effects}

We incorporate the $t\bar t$ threshold effects into the Green function
by evaluating it with the QCD potential~\cite{Fadin:1987wz,Fadin:1988fn}.
The amplitude with the threshold effects is expressed
as~\cite{Melnikov:1994jb} 
\begin{align}
 M_{t,\{\lambda_i\}}^{\rm thr} = {\mathcal A}^{J=0}_{t,\{\lambda_i\}} +
 {\mathcal B}_{t,\{\lambda_i\}} G(\vec0;{\mathcal E})
 + {\mathcal A}^{J>0}_{t,\{\lambda_i\}}(\theta) ,
 \label{eq:thr}
\end{align}
where we define ${\mathcal A}^{J>0}_t(\theta) = {\mathcal A}_t(\theta) -
{\mathcal A}_t^{J=0}$ and ${\mathcal E}=E+i\Gamma_t$ with the top-quark
decay width, $\Gamma_t$. 
The Green function is defined by the following Schr\"odinger equation:
\begin{align}
 \left[\left\{-\frac{\nabla^2}{m_t} +
 V(r) \right\} - {\mathcal E} \right] 
 G(\vec r;{\mathcal E}) = \delta^3(\vec r) ,
 \label{eq:sch}
\end{align}
where $V(r)$ is the QCD potential.
For the $t\bar t$ system, we can utilize the perturbatively-calculated
potential.
The real part of the Green function at $\vec r=\vec0$ is known to be
divergent, thus has to be renormalized.
We adopt the $\overline{\rm MS}$ renormalization scheme in dimensional
regularization~\cite{Beneke:1999ff,Beneke:1999zr,Hoang:2001mm}.
An artificial scale $\mu$ is introduced to the renormalized Green
function.
By matching with the one-loop amplitude, the amplitude is finally
expressed as 
\begin{align}
 M_{t,\{\lambda_i\}}^{\rm match} & = M_{t,\{\lambda_i\}} + {\mathcal
 B}_{t,\{\lambda_i\}}\left[ G(\vec0;{\mathcal E}) - G^{(0)}(\vec0;E) \right]
 .
\end{align}

Before moving to the numerical evaluation, we discuss the order of the
corrections in the non-relativistic QCD formalism. 
Taking $v$ and $\alpha_s$ as the expansion parameters, the leading-order
contribution is the ${\mathcal A}_t^{J=0}$ term which is constant, and
the ${\mathcal B}_tG(\vec0;{\mathcal E})$ term is at the next-to-leading
order (NLO).
There is another NLO term in the two-loop amplitude, which is an
${\mathcal O}(\alpha_s)$ correction to ${\mathcal A}_t$. 
However, this has not been calculated yet for the massive-quark
contribution.
Indeed, this term is required for the consistent calculation of the
threshold corrections up to NLO in order that the scale dependence of
the real part of the Green function is canceled with the ${\mathcal
O}(\alpha_s)$ term of ${\mathcal A}_t$~\cite{KY}.
In our calculation, we do not include the ${\mathcal O}(\alpha_s)$
${\mathcal A}_t$-term, thus the scale dependence remains in the
threshold amplitude.
We treat it as an uncertainty of our calculation.

Since the leading contribution to the squared amplitude is the absolute
square of the sum of $M_q$ and the ${\mathcal A}_t$ term where both are
independent of energy, the uncertainty of ${\mathcal A}_t$ term mainly
affects the overall normalization of the diphoton mass spectrum. 
On the other hand, some of the NNLO corrections improve the description
of the $t\bar t$ resonances.
Therefore, for the sake of a precise and stable prediction of the
resonance structure, it is worthwhile to include the available NNLO
corrections even though we cannot reach the full NLO accuracy.
The known corrections are (1) the NLO correction to the Green function,
(2) the ${\mathcal O}(\alpha_s)$ correction to ${\mathcal B}_t$, and (3)
the ${\mathcal O}(\alpha_s)$ correction to $\Gamma_t$.
First, the NLO correction to the Green function is incorporated by
solving the Schr\"odinger equation with the NLO QCD
potential~\cite{Fischler:1977yf,Billoire:1979ih} given by
\begin{align}
 V(r) = - C_F \frac{\alpha_s(\mu_B)}{r} \left[ 1 +
 \frac{\alpha_s}{4\pi} \left\{ 2\beta_0\left[ \ln{(\mu_B r)} + \gamma_E
 \right] + a_1 \right\} \right], 
\end{align}
where $\beta_0=11/3\cdot C_A-2/3\cdot n_f$ and $a_1=31/9\cdot
C_A-10/9\cdot n_f$ with $C_F=4/3$ and $C_A=3$.
We will show later that evaluating the Green function beyond LO is
crucial for the reliable prediction.
Second, the ${\mathcal O}(\alpha_s)$ correction to ${\mathcal
B}_t$ can be derived from the ${\mathcal O}(\alpha_s)$
hard-vertex corrections to the $gg\to t\bar t$ and $t\bar
t\to\gamma\gamma$ processes. 
The hard-vertex factor to the $gg\to t\bar t$ cross-section in the
color-singlet channel reads $1+(\alpha_s/\pi)h_1$
with~\cite{Petrelli:1997ge,Hagiwara:2008df,Kiyo:2008bv}
\begin{align}
 h_1 = C_F \left( - 5 + \frac{\pi^2}{4} \right)+ C_A \left( 1 +
 \frac{\pi^2}{12} \right) + \beta_0\ln{\left(\frac{\mu_R}{2m_t}\right)} ,
\label{eq:h1}
\end{align}
where $\mu_R$ is the renormalization scale of $\alpha_s$.
The corresponding factor for $t\bar t\to\gamma\gamma$ reads only the
first term of Eq.~(\ref{eq:h1}).
By using them, ${\mathcal B}_t$ with the ${\mathcal O}(\alpha_s)$
correction is given as ${\mathcal B}_t = {\mathcal B}^{(0)}_t[ 1 +
(\alpha_s/\pi) b_1 ]$ with 
\begin{align}
b_1 = C_F \left( - 5 +
 \frac{\pi^2}{4} \right) + \frac{C_A}{2} \left( 1 + \frac{\pi^2}{12}
 \right) + \frac{\beta_0}{2} \ln{\left(\frac{\mu_R}{2m_t}\right)} .
\end{align}
Finally, the ${\mathcal O}(\alpha_s)$ correction to $\Gamma_t$ has been
calculated in Refs.~\cite{Jezabek:1988iv,Czarnecki:1990kv,Li:1990qf}.
However, we treat $\Gamma_t$ as an input parameter in our study.
Identification and derivation of the remaining NNLO corrections are
beyond the scope of this paper.

\section{Numerical results}

\begin{figure}[t]
 \begin{center}
  \includegraphics[width=.49\textwidth]{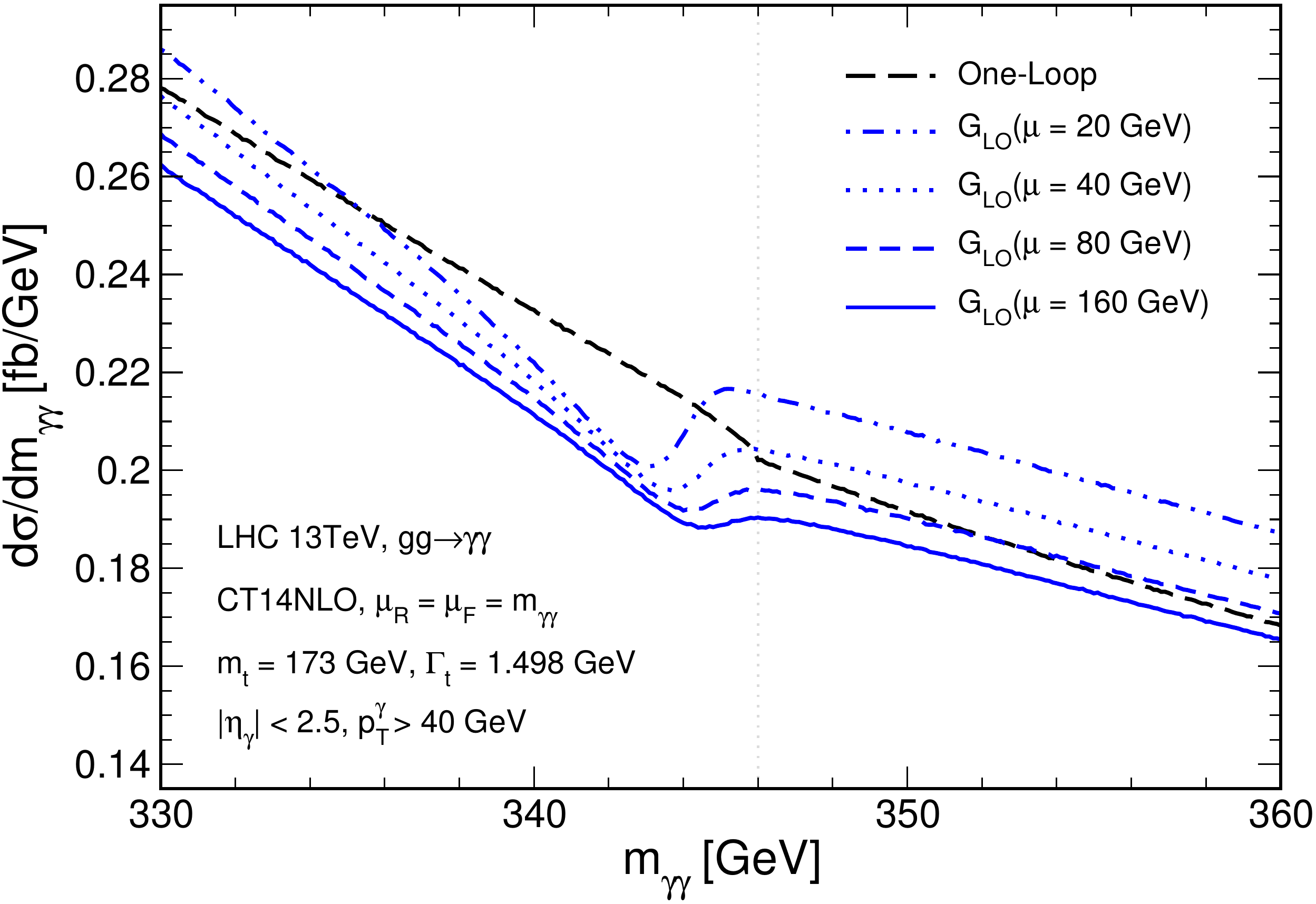} 
  \includegraphics[width=.49\textwidth]{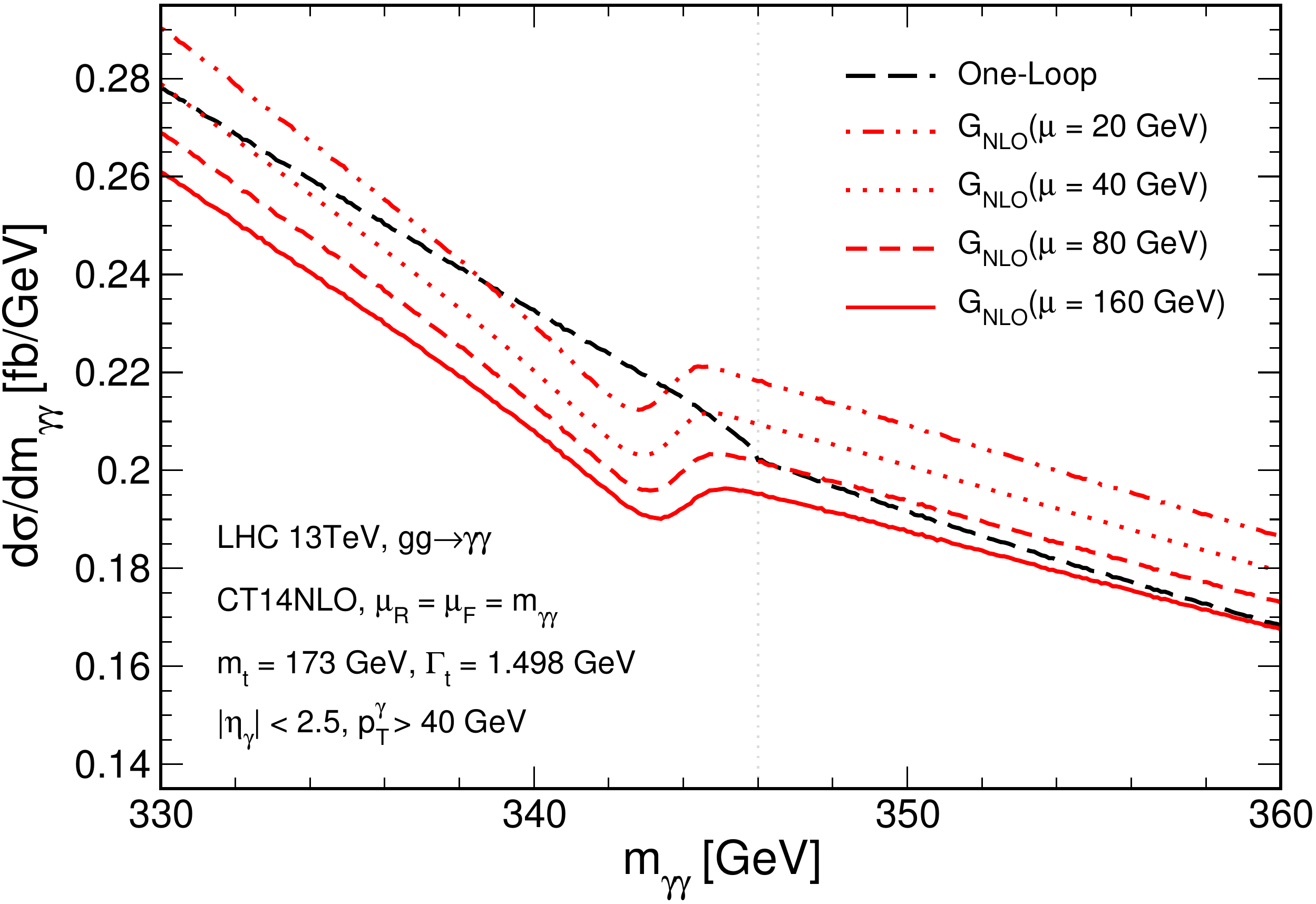}
  \caption{$gg\to\gamma\gamma$ cross sections near the $t\bar t$
  threshold at the LHC 13~TeV.
  Left panel is for the LO Green function, and right panel is for the
  NLO Green function.
  }\label{fig:LHC13}
 \end{center}
\end{figure}

We present numerical studies for the $gg\to\gamma\gamma$ amplitude
as well as the cross sections at the LHC.
In Fig.~\ref{fig:LHC13}, we plot $d\sigma/dm_{\gamma\gamma}$ via
$gg\to\gamma\gamma$ for the LHC 13~TeV with kinematical cuts of
$|\eta_\gamma|<2.5$ and $p_{T}^\gamma>40$~GeV~\cite{Aaboud:2016tru}.
Both the massive- and massless-quark loops are included.
We use the {\tt CT14NLO} gluon distribution
function~\cite{Dulat:2015mca}, and take the renormalization and
factorization scales as $\mu_R=\mu_F=m_{\gamma\gamma}$.
The Green function is evaluated by numerically solving Eq.~(\ref{eq:sch})
with the LO or NLO QCD potential following the method described in
Ref.~\cite{Kiyo:2010jm}.
The scale of $\alpha_s$ in the QCD potential is taken as the same as
the renormalization scale $\mu$ of the Green function, which we vary
from 20~GeV to 160~GeV.
The result with the one-loop amplitude is also plotted for comparison.
In the plots, we observe that the distributions show a characteristic
structure near $m_{\gamma\gamma}\simeq2m_t=346$~GeV; it shows a dip
and then a small bump below the threshold~\cite{Chway:2015lzg}.
We find that, if we employ the LO Green function, the shape of the
distribution changes by the scale choice.
In contrast, by using the NLO Green function
the shape of the distribution is quite stable apart from the overall
normalization. 
The positions of the dip and the bump are shifted by the choice of
$\mu$ by around 0.6~GeV.
A relatively large uncertainty appears as the overall size of the cross
section, which amounts to about 10\%.
This uncertainty originates mainly from the lack of the ${\cal
O}(\alpha_s)$ correction in the ${\mathcal A}_t$ term.
We note that there exists another source of the uncertainty for the
overall normalization, which is the scale choice of $\mu_R$ and $\mu_F$. 
For the LHC 13~TeV, changing these scales from $m_{\gamma\gamma}/2$ 
to $2m_{\gamma\gamma}$ varies the cross section by about 20\%. 

\begin{figure}[t]
 \begin{center}
  \includegraphics[width=.75\textwidth]{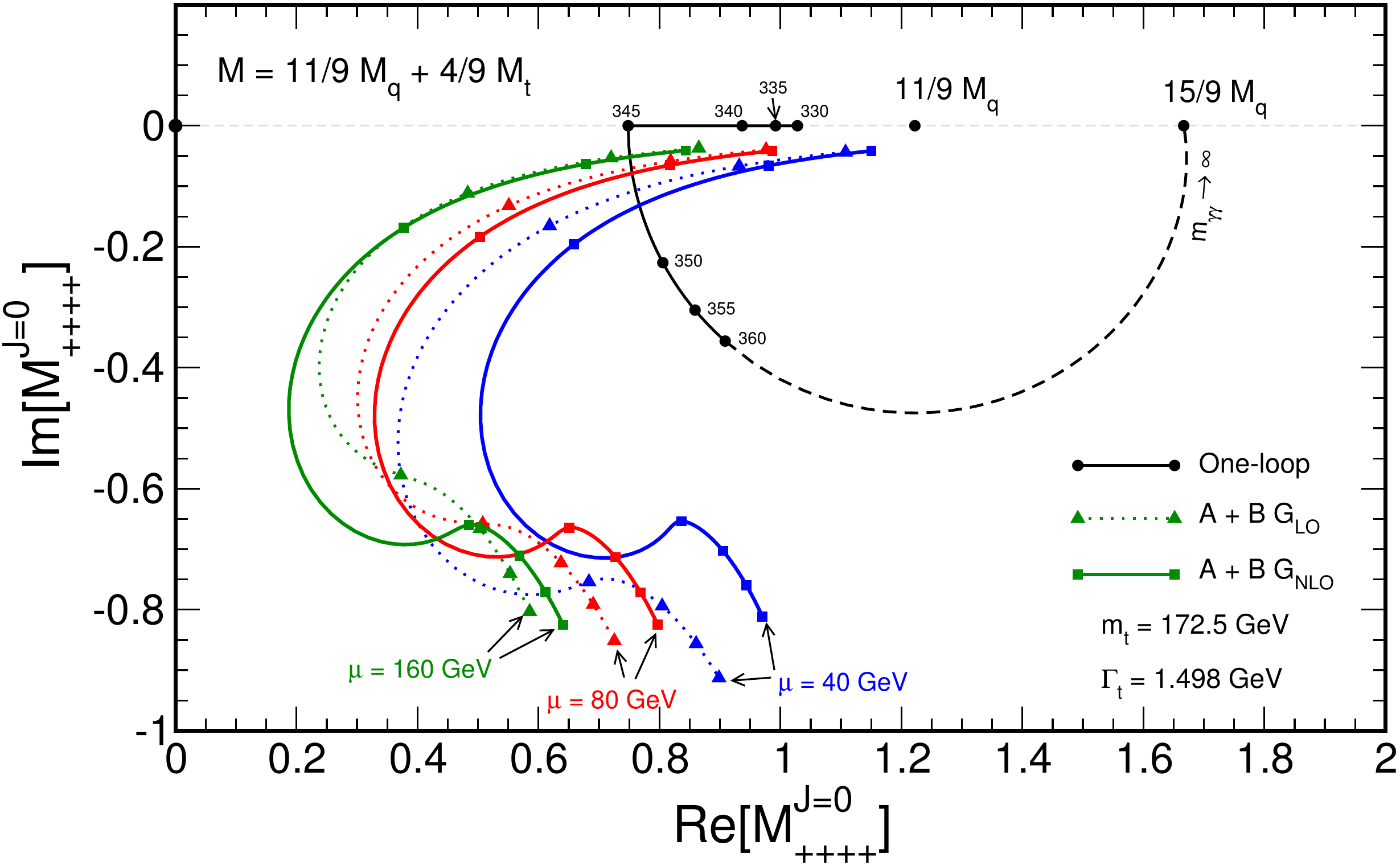}
  \caption{$gg\to\gamma\gamma$ amplitudes for $\{\lambda_i\}=++++$ in
  the $J=0$ channel for $m_{\gamma\gamma}=330$~GeV to 360~GeV with
  points in a 5-GeV step.
  For the illustrative convenience, we set $m_t=172.5$~GeV in this
  plot. } \label{fig:Amp}
 \end{center}
\end{figure}

For a better understanding of the behavior of the cross section, 
we plot in Fig.~\ref{fig:Amp} the $gg\to\gamma\gamma$ amplitudes in a
complex plane for $\{\lambda_i\}=++++$ in the $J=0$ channel, by
varying $m_{\gamma\gamma}$ from 330~GeV to 360~GeV. 
The massless-quark-loop amplitude gives a constant contribution,
$11/9\cdot M_{q,++++}^{J=0}=11/9$.
The total amplitude $M^{J=0}_{++++} = 11/9\cdot M^{J=0}_{q,++++} +
4/9\cdot M^{J=0}_{t,++++}$ with the one-loop-level $M^{J=0}_{t,++++}$ is
drawn in the black line.
Below the threshold, the two amplitudes, $M_q$ and $M_t$, are pure real,
and their relative sign is negative.
Therefore, there is a destructive interference, and the total amplitude
goes toward the origin by increasing $m_{\gamma\gamma}$ until the
threshold.
Above the threshold, the amplitude gains an imaginary part and the
real part tends to increase along with $m_{\gamma\gamma}$.
At the high-energy limit, where the top quark can be assumed to be
massless, the imaginary part goes to zero and the total amplitude
arrives at $M^{J=0}_{++++}=15/9$.\footnote{%
Interference effects with $s$-channel resonant diagrams have been
studied in
Refs.~\cite{deFlorian:2013psa,Dicus:1987fk,Martin:2012xc,Jung:2015sna,%
Jung:2015etr,Djouadi:2016ack}
}
The amplitude with the threshold corrections calculated with the NLO
(LO) Green functions are plotted in colored solid (dotted) lines for
$\mu=40$, 80 and 160~GeV.
The imaginary part of the amplitude is non-zero even below the
threshold, which comes from the finiteness of $\Gamma_t$.
The size of the imaginary part increases rapidly above
$m_{\gamma\gamma}=340$~GeV with showing a resonance-like curve just
below the threshold.
The scale dependence of the Green function originates from the two
sources, one in the QCD potential and the other from the
real-part renormalization.
For the NLO Green function, the former is well suppressed and the latter
affects only the real part of the amplitude by a constant for any
$m_{\gamma\gamma}$.
For the LO Green function, both effects are large and the amplitude
shows a complicated scale dependence.
Especially, there remains a scale dependence in the imaginary part of
the amplitude.
This explains the reason that the shape of the invariant-mass
distribution is stable by using the NLO Green function in contrast to
the LO Green function.
Although the uncertainty in the real part of the amplitude is
significant, it leads only the 10\% level uncertainty to the cross
section, due to the presence of the large imaginary part and the
light-quark-loop contribution.

\begin{figure}[t]
 \begin{center}
  \includegraphics[width=.6\textwidth]{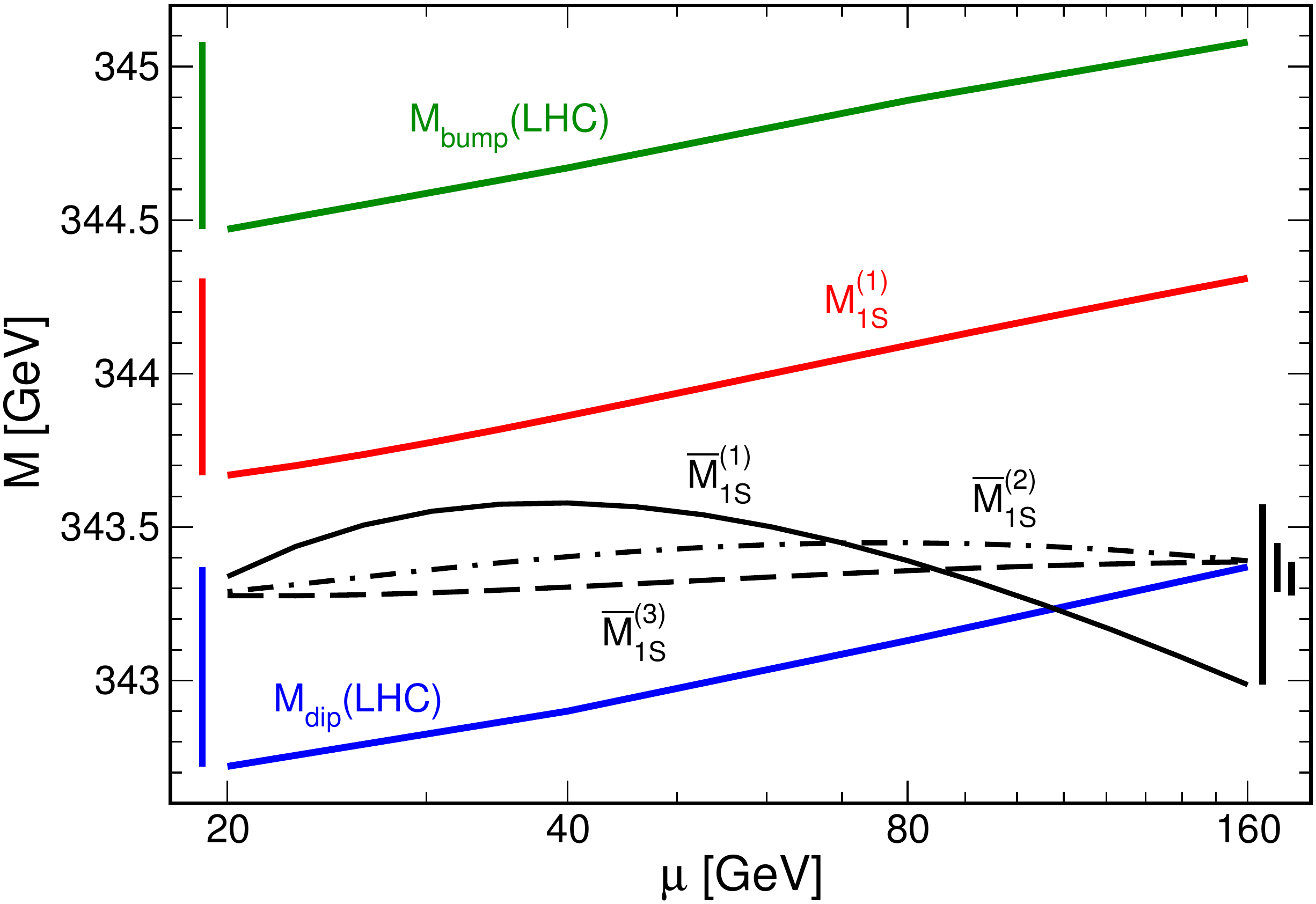} \\
  \caption{Scale dependence of the dip and bump positions in the
  diphoton mass spectrum at the LHC evaluated with the NLO Green
  function, and the energy level of the $1S$ toponium evaluated in the
  pole-mass scheme at NLO as well as those in the $\overline{\rm 
  MS}$-mass scheme up to N$^{3}$LO.
  $m_t^{\rm pole}=173$~GeV or $\overline{m}_t = 163$~GeV is
  used. }\label{fig:M1S}
 \end{center}
\end{figure}
In Fig.~\ref{fig:M1S}, we show the scale dependence of the dip and bump
positions, $M_{\rm dip}$ and $M_{\rm bump}$, respectively, in the
diphoton mass spectrum at the LHC evaluated with the NLO Green
function.
In addition, we plot the $1S$ energy-level of the $t\bar t$ bound-state
(toponium) at NLO [${\cal O}(\alpha_s^3m_t)$], $M^{(1)}_{1S}$, which is
in good approximation the resonance peak position in the NLO Green
function. 
We find the scale variation of the Green function affects the difference
of the two mass scales, $M_{\rm dip}$ and $M_{1S}$~(and also, $M_{\rm
bump}$ and $M_{1S}$), by only around 20~MeV~(40~MeV). 
This indicates that the connection of the dip~(bump) position and the
$1S$ resonance mass is sufficiently solid under uncertainties of the
Green function. 
The toponium energy-levels have been calculated up to ${\cal
O}(\alpha^5_sm_t)$~\cite{Penin:2002zv,Beneke:2005hg} in non-relativistic
QCD, and it is well-known that the prediction becomes significantly
accurate when it is expressed in terms of the short-distance mass to
cancel the renormalon ambiguity. 
By using the ${\cal O}(\alpha^5_sm_t)$ formula for the spin-singlet
case~\cite{Penin:2002zv,Beneke:2005hg} and the $\overline{\rm MS}$ mass
with $\overline{m}_t=163$~GeV, we also plot the $1S$ energy-level,
$\overline{M}^{(n)}_{1S}$, at N$^{n}$LO up to $n=3$ in
Fig.~\ref{fig:M1S}.
It can be seen that the convergency is good, and the scale uncertainty
is reduced to around 100~MeV or below.\footnote{%
In Ref.~\cite{Kiyo:2015ooa}, the scale variation is examined for a range
from 80 to 320~GeV, and the uncertainty is claimed to be about
40~MeV.}
By combining these arguments, the dip and bump positions can be
accurately predicted by including higher-order corrections with the
short-distance mass.
More detailed studies will be presented in a future publication.

\section{Top-quark mass from the diphoton mass spectrum}

\begin{figure}[t]
 \begin{center}
  \includegraphics[width=.492\textwidth]{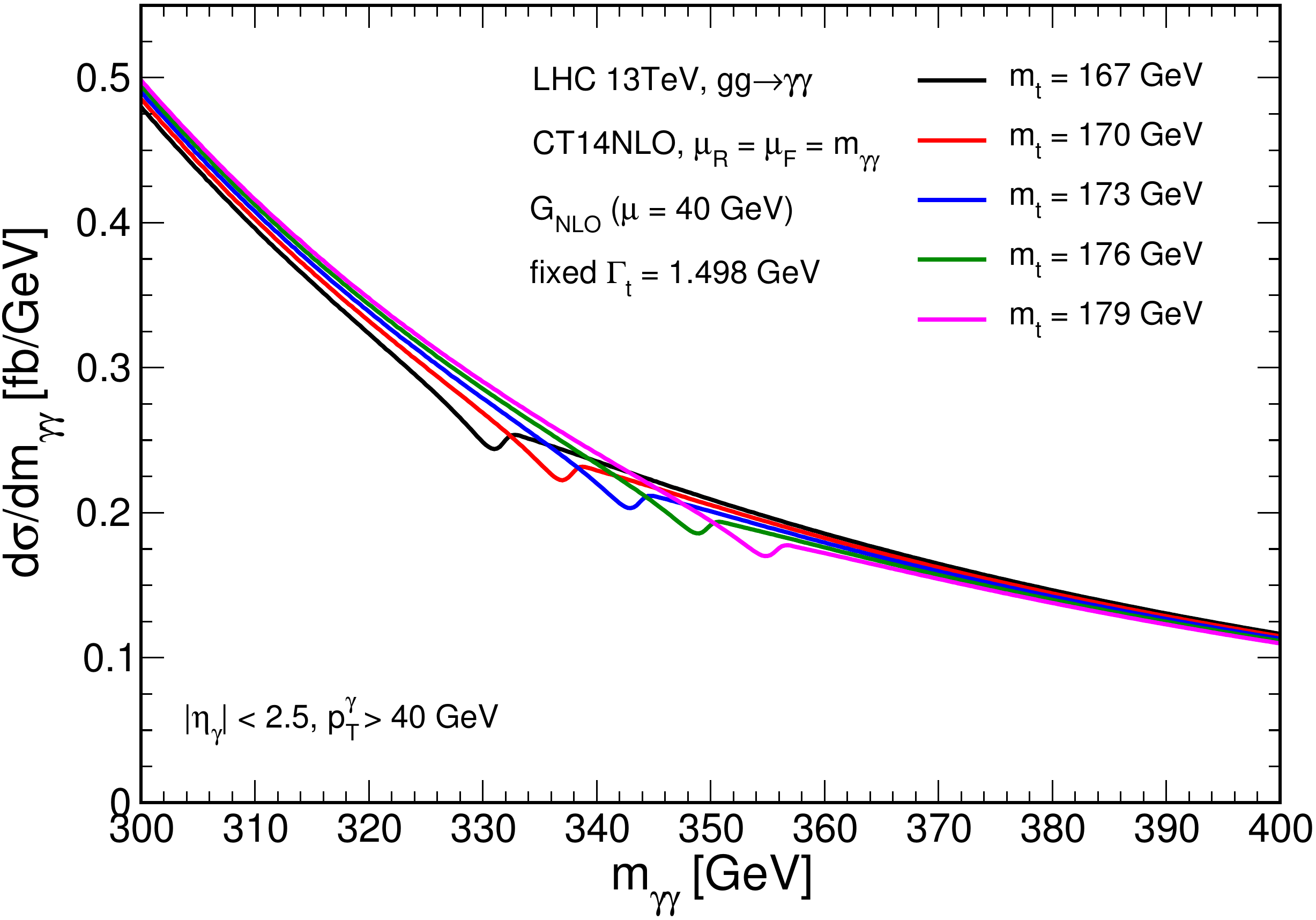}
  \includegraphics[width=.48\textwidth]{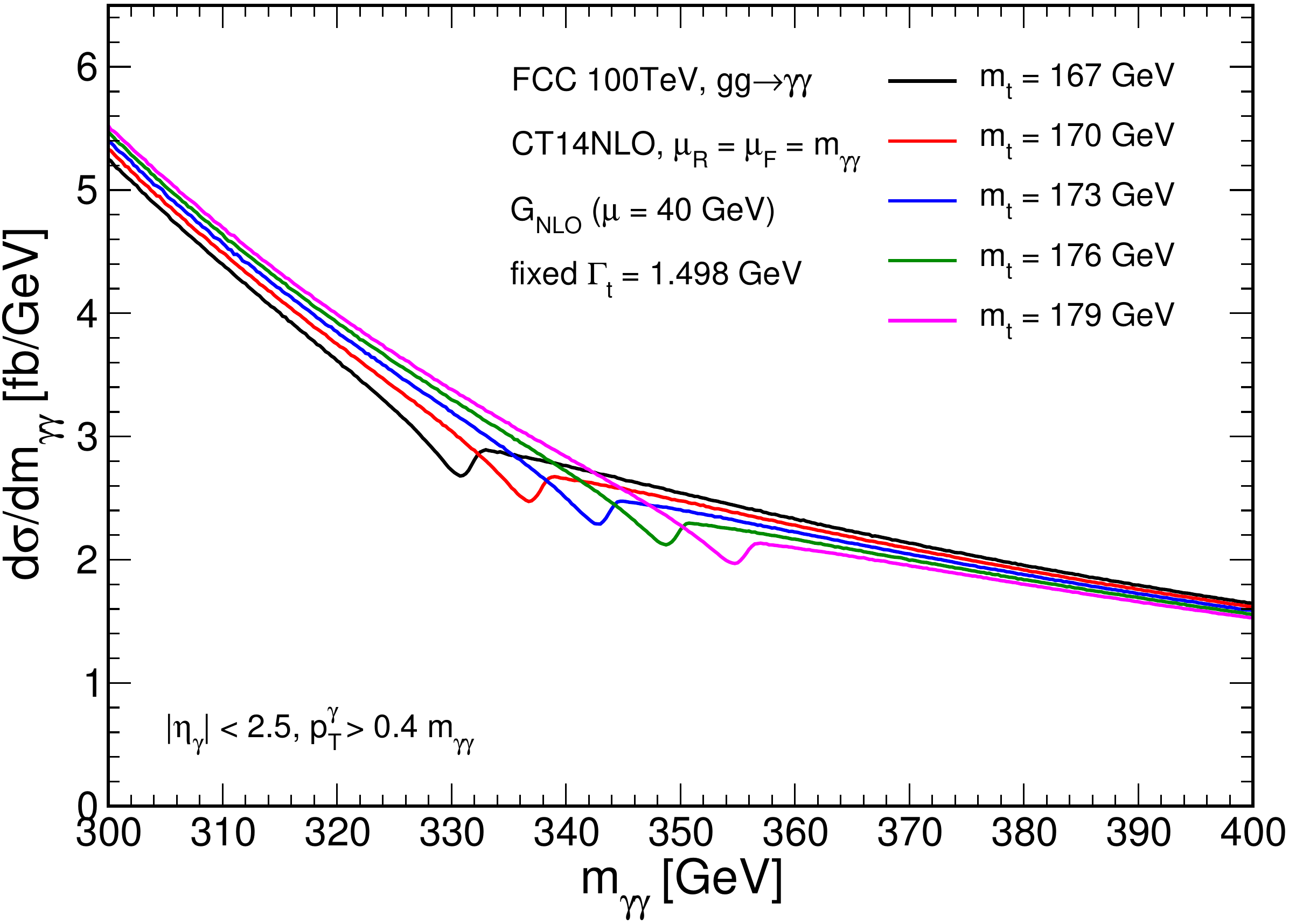}
  \caption{$gg\to\gamma\gamma$ differential cross sections for different
  $m_t$.
  Left panel is for the LHC 13~TeV and right panel is for the FCC
  100~TeV. }\label{fig:Temp}
 \end{center}
\end{figure}

We propose to use the diphoton mass spectrum near the $t\bar t$
threshold for a precise determination of the top-quark mass in
hadron-collider experiments.
Fig.~\ref{fig:Temp} shows the diphoton mass spectra via
$gg\to\gamma\gamma$ with different values of $m_t$ for the LHC at
$\sqrt{s}=13$~TeV (top panel) and the proposed future circular collider
(FCC) at
$\sqrt{s}=100$~TeV~\cite{Hinchliffe:2015qma,Arkani-Hamed:2015vfh,%
Mangano:2016jyj}~(bottom panel).
We utilize the NLO Green function with $\mu=40$~GeV.
$\Gamma_t=1.498$~GeV is fixed for any $m_t$.
The setup for the gluon distribution function and acceptance cuts
is same as that for Fig.~\ref{fig:LHC13}.
An additional cut $p_T^\gamma>0.4m_{\gamma\gamma}$ is applied for the
FCC case which enhances the selection efficiency of the $J=0$ partial-wave
contribution.
One can clearly see in Fig.~\ref{fig:Temp} that the bump position shifts
in proportion to $m_t$.
Consequently, we can extract $m_t$ from the diphoton mass spectrum.
Since a photon is a clean object and not directly affected by
final-state QCD interactions, this measurement would be quite
transparent experimentally and theoretically. 
Especially, systematic errors of photon momentum reconstruction are much
smaller than those of jet momentum which are the major source of the
systematic error in the current $m_t$ measurement.
These virtues are shared with leptonic-observable methods proposed in
Refs.~\cite{Kharchilava:1999yj,Kawabataa:2014osa,Frixione:2014ala}.

In order to estimate the sensitivity of the method, we perform
pseudo-experiments assuming the LHC 13~TeV with 3~ab$^{-1}$ data, 
and the FCC 100~TeV with 1~ab$^{-1}$ and 10~ab$^{-1}$ data.
We prepare event samples of the {\it signal} $gg\to\gamma\gamma$ events
and the {\it background} events by other sources for the range
$m_{\gamma\gamma}=[300,400]$~GeV with applying the above acceptance
cuts. 
The signal events are generated based on the predicted distribution
assuming $m_t^{\rm true}=173$~GeV. 
The background events are generated by {\tt Diphox}~\cite{Binoth:1999qq}
at LO with $q\bar q\to\gamma\gamma$, one-direct--one-fragmentation, and 
two-fragmentation contributions. 
The total number of events is fixed by using the observed data to
take into account detector efficiency and a $K$-factor from
higher-order corrections.
We read off a corresponding correction factor of $C\simeq1.2$ from the
LHC 13~TeV diphoton analysis by the ATLAS
Collaboration~\cite{ATLAS:2016eeo}. 
For simplicity we apply the same $C$ for the FCC case. 
The signal-to-background ratio, which is crucial to the mass
sensitivity, is subject to theoretical uncertainties of the
cross-section calculations, such as the choice of scales $\mu_R$, 
$\mu_F$ and $\mu$, uncalculated higher-order corrections, and also a
definition of isolated photons~\cite{Frixione:1998jh}.
Based on the LO calculations for both the signal and background
processes, the ratio is estimated to be 10\% at the LHC 13~TeV and 30\%
at the FCC 100~TeV. 
On the other hand, the ratio is estimated to be 5\% at the LHC where the
QCD NNLO corrections are included in the background
calculation~\cite{Catani:2011qz,Campbell:2016yrh}, while 10\% at the
FCC where the QCD NLO corrections are included in the background
calculation~\cite{Mangano:2016jyj}. 
We note that a recent study in Ref.~\cite{Campbell:2016yrh} indicates
that the ratios become closer to the LO estimates when the NLO
corrections are included additionally to the signal process.
Considering these estimations, we take the ratio to be 5\% to 10\%
at the LHC, while 10\% to 30\% at the FCC in this study.

The sample $m_{\gamma\gamma}$ distributions are fitted with the sum of
the signal prediction which depends on $m_t$ plus an analytic smooth
function for the background, taking the signal-to-background ratio
as a fitted parameter.
The background function is taken as $(1-x^{1/3})^a$ where
$x=m_{\gamma\gamma}/\sqrt{s}$ and $a$ is a parameter to be fitted.
Notice that our fitting procedure does not rely on the value of the
signal-to-background ratio nor the accurate prediction of the
background shape.
We perform least-squares fits to the binned $m_{\gamma\gamma}$
distribution in the interval [300, 400]~GeV with the bin width of 1~GeV.
By repeating the pseudo-experiment, we obtain the expected statistical
error $\Delta m_{t}$ from the distribution of the fitted $m_t$.
For the LHC 3~ab$^{-1}$, the obtained $m_t$ distribution is not
Gaussian, while it has a peak at $m_t=m_t^{\rm true}$. 
We approximate the distribution as Gaussian and obtain
$\Delta m_t\simeq2$~GeV to 3~GeV for the signal ratio 10\% to 5\%.
For the FCC, by assuming the signal-to-background ratio to be 30\%, the
distribution behaves as Gaussian and we obtain $\Delta m_t=0.2$~GeV
(0.06~GeV) for 1~ab$^{-1}$~(10~ab$^{-1}$). 
When the ratio is assumed to be 10\%, we obtain $\Delta m_t=0.6$~GeV
(0.2~GeV) for 1~ab$^{-1}$~(10~ab$^{-1}$).
We find that the correlation between two fitted parameters, $m_t$ and
the signal-to-background ratio, is weak. 

Before closing, we present several comments.
The systematic error of photon energy scale is about
0.5\%~\cite{Aad:2014nim} in the ATLAS detector and about
0.3\%~\cite{Khachatryan:2015iwa} in the CMS detector.
Thus we naively expect the systematic error of $\delta m_t^{\rm
sys.}\lesssim1$~GeV at the future LHC measurement.
For more realistic estimation at the LHC as well as at the FCC,
simulation studies with detailed detector performance are required.
Beyond the one-loop level, the mass renormalization scheme becomes
explicit.
With the signal distribution expressed in terms of theoretically
well-defined masses, the top-quark $\overline{\rm MS}$ mass can be
extracted directly from the diphoton mass spectrum.
Measuring the short-distance mass from the resonance structure is
conceptually equivalent with the threshold scan method in $e^+e^-\to
t\bar t$.
In the $e^+e^-$ case, the threshold production cross-section is
established up to N$^3$LO in non-relativistic
QCD~\cite{Beneke:2013jia,Beneke:2015kwa}. 
In the diphoton case at hadron colliders, only the one-loop
$gg\to\gamma\gamma$ amplitude has been known, and thus the NLO
calculation has not been completed yet.
To complete, one requires the two-loop $gg\to\gamma\gamma$, one-loop
$gg\to\gamma\gamma g$ and $gq\to\gamma\gamma q$ amplitudes.
In the one-gluon emission processes, corrections via an initial-state
gluon emission, color-octet $t\bar t$ effects, and an ultrasoft gluon
emission from the on-shell $t\bar t$ state appear.
These corrections can be sizable because of the large partonic
luminosity of the color-octet gluons.
Investigations of these effects are left for future works. 
However, we expect that these would not severely spoil the
characteristic shape of the spectrum in the resonance region, because
the initial-state radiation does not affect the bound-state formation
and the color-octet $t\bar t$ Green function is known to have a smooth
slope in the resonance region. 
Finally, it might be possible to determine $\Gamma_t$ simultaneously
with $m_t$ at the FCC.

\section{Conclusions}

To conclude, we have studied the $gg\to\gamma\gamma$ amplitude with
the $t\bar t$ bound-state effects near their mass threshold by
collecting the available higher-order corrections in non-relativistic
QCD. 
We have predicted a characteristic structure in the diphoton mass
spectrum near the threshold whose shape is stable under the scale
uncertainty, while the overall normalization has an uncertainty of 10\%
level due to the lack of the two-loop amplitude.
We have proposed a new method to determine $m_t$ from the diphoton
mass spectrum at the LHC and the FCC.
We have shown that the estimated statistical errors are fairly small at
the FCC, which deserves further realistic experimental studies and also
motivates to calculate higher-order corrections in theory.

\begin{acknowledgments}
We are grateful to Yukinari~Sumino and Yuichiro~Kiyo for valuable
 discussions and encouragements.
We also thank Hyung Do~Kim and Michihisa Takeuchi for useful
 discussions.
The research of S.K.\ was supported by Basic Science Research Program
through the National Research Foundation of Korea (NRF) funded by the
Ministry of Science, ICT and Future Planning (Grant No.\
NRF-2014R1A2A1A11052687). 
\end{acknowledgments}


\end{document}